\documentclass[onecolumn,superscriptaddress,floatfix,preprintnumbers,amssymb ,amsmath]{revtex4}
\usepackage{graphicx}
\usepackage{dcolumn}
\usepackage{bm}
\usepackage[latin1]{inputenc}
\usepackage[mathscr]{eucal}
\usepackage{epsfig}

\begin{document}
\title{Work, free energy and dissipation in voltage driven single-electron transitions}

\author{J. P. Pekola}
\author{O.-P. Saira}
\affiliation{Low Temperature Laboratory (OVLL), Aalto University School of Science, P.O. Box 13500, 00076 Aalto, Finland}

\begin{abstract}
We apply the general procedure presented by Jarzynski in \cite{j1} to the problem of dissipation in a voltage-driven single-electron box. We obtain the expression of dissipated work, and find its relation to the dissipation $Q$ obtained in \cite{ap11}. We show that the two quantities are identical in common gate protocols where the system makes a transition for sure.
\end{abstract}
\date{\today}
\maketitle

\section{Introduction}
\label{intro}
\begin{figure}
    \includegraphics[width=10cm]{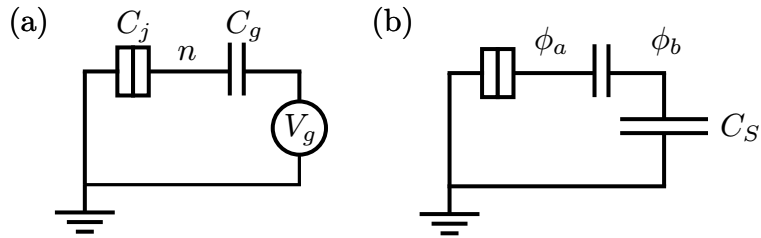}
    \caption{(a) The single-electron box with a voltage source for controlling the gate charge. (b) The voltage source represented by a large capacitor.}
    \label{fig:box}
\end{figure}

The non-equilibrium fluctuation relations \cite{bochkov81,jarzynski97,crooks99} yield important results for the distribution of dissipation in experiments where external control parameters are varied non-adiabatically~\cite{liphardt02,collin05,alemany11}. The validity of these relations rests on appropriate expressions for work and free-energy. This issue was highlighted in a scholarly manner in \cite{j1}. The presented general treatment allows one to apply it to a specific system, which in the current case is a single-electron box (SEB) \cite{mb,sac}, see Fig.~\ref{fig:box}(a). Single-electron transitions \cite{averin86} yield a suitable test-bench of fluctuation relations, where reliable statistics can be collected under stable experimental conditions \cite{ops}. In a SEB, a tunnel contact, schematically a split box on the left in Fig.~\ref{fig:box}(a) admits electrons to enter or leave the island in the middle at rates determined by standard tunneling expressions. The characteristic electrostatic energy of the box can be made large (small size, low temperature) such that the number of electrons on it is limited to only two neighbouring values in a given control parameter (gate) interval. Our analysis is, however, not restricted to such two-state dynamics.

The quantity $Q$ introduced in Ref.~\cite{ap11} is the dissipated energy in the tunneling transitions only, and thus it represents the heat generated. In the present paper we demonstrate mathematically that the dissipated work, $W-\Delta F$, relevant in fluctuation relations~\cite{j1,jarzynski97,crooks99}, consists of the sum of $Q$ and the electrostatic energy stored in the circuit during the driven evolution.

\section{Work and free-energy in a single-electron box}
We discuss the hamiltonian $H$ of the SEB, yielding the work $W$ in a given protocol of the external drive voltage, and the equilibrium free-energy $F$ at a given voltage bias, along the general prescription given in \cite{j1}.

The "bare" hamiltonian $H_0$ of the SEB is given by the bare electrostatic energy of the capacitors in the circuit,
\begin{eqnarray} \label{box1}
H_0 = \frac{Q_j^2}{2C_j}+\frac{Q_g^2}{2C_g},
\end{eqnarray}
where $C_j$ is the capacitance of the tunnel junction, and $C_g$ is that of the gate, see Fig. \ref{fig:box}(a); $Q_j$ is the charge at the junction and $Q_g$ is that of the gate capacitor.

The generalized force of the SEB is the gate voltage $V_g$, and the generalized coordinate is its charge $Q_g$. With the help of these, we can express the transformed hamiltonian $H$ as
\begin{eqnarray} \label{box3}
H=H_0-Q_gV_g.
\end{eqnarray}
We give a proper justification of this basic transformation in Section \ref{sec:lagrange}. We next write down $H$ as a function of the integer number $n$ of electrons on the island and the normalized gate voltage $n_g=-C_gV_g/e$. The gate charge is given by $Q_g=C_g(V_g-\dot \phi_a)$. Here $\dot\phi_a$ is the island potential
\begin{eqnarray} \label{box2}
\dot\phi_a=\frac{e}{C_\Sigma}(n-n_g),
\end{eqnarray}
and $C_\Sigma = C_j + C_g$ is the total capacitance of the SEB.
By elementary analysis we then have
\begin{eqnarray} \label{box4}
H(n,n_g) = E_C n^2 -2E_C n n_g -\frac{e^2}{2}(C_g^{-1}-C_\Sigma^{-1})n_g^2,
\end{eqnarray}
where $E_C=e^2/(2C_\Sigma)$ is the charging energy of the SEB.

The work performed in a driven process can be written in general as $W=\int \frac{\partial H}{\partial \lambda}\dot \lambda dt$, where $\lambda$ is the control parameter and the dot refers to the derivative with respect to time $t$ \cite{j1}. Thus the work for our system reads
\begin{eqnarray} \label{box6b}
W = \int \frac{\partial H}{\partial V_g} dV_g,
\end{eqnarray}
where the integral is extended over the gate excursion between the end points.
Direct differentiation of Eq. (\ref{box4}) yields $\frac{\partial H}{\partial V_g}=-Q_g$ and the
proper "thermodynamic" work is
\begin{eqnarray} \label{box6}
W = -\int Q_g dV_g
\end{eqnarray}
in full analogy to the result in \cite{j1}.
The "standard" expression of work is, on the other hand, the integral of the force along the position, i.e.,
\begin{eqnarray} \label{box5}
W_0 = \int V_g dQ_g.
\end{eqnarray}
As discussed in Ref. \cite{j1}, the two expressions (\ref{box6}) and (\ref{box5}) are identical for a cyclic trajectory where the perturbation $V_g$ is turned on and then off, as can be verified by elementary partial integration.

With the notations we have used above, we find
\begin{eqnarray} \label{box7}
W =-2E_C\int_{n_{g,A}}^{n_{g,B}} n dn_g -\frac{e^2}{2}(C_g^{-1}-C_\Sigma^{-1})(n_{g,B}^2-n_{g,A}^2),
\end{eqnarray}
where $n_{g,A}$ and $n_{g,B}$ are the gate positions in the beginning and at the end, respectively.

Next we find the change in free energy, again according to the standard procedure \cite{j1}. For completeness, we allow the system to occupy any integer valued charge state $n$ although it is exponentially unlikely for the system to visit a state for which $|n - n_g| \gg 1/\sqrt{\beta E_C}$, where $\beta$ is the inverse temperature of the bath.  For writing down the partition function $Z(n_g)$, we first regroup the Hamiltonian as $H = E_C(n - n_g)^2 - e^2/(2 C_g) n_g^2$, and then perform the summation over possible $n$ states as
\begin{eqnarray} \label{box8}
Z(n_g)= \sum_{n} e^{-\beta H(n,n_g)} = \exp\big(\frac{\beta e^2  n_g^2}{2C_g}\big) \sum_{n} e^{-\beta E_C (n - n_g)^2}.
\end{eqnarray}
The remaining infinite summation defines a special function which we will refer to as $R_{\beta E_c} (n_g)$. The function is periodic in $n_g$ with a period of 1, which is a manifestation of the fact that there is no absolute reference point for the charge number $n$.

The change of free energy for the given gate sweep $n_{g,A}\rightarrow n_{g,B}$ is then
\begin{eqnarray} \label{box10}
\Delta F&& = F(n_{g,B})-F(n_{g,A}) = -\beta^{-1} \ln \frac{Z(n_{g,B})}{Z(n_{g,A})} \nonumber \\&&= -\frac{e^2}{2C_g} \left(n_{g,B}^2 - n_{g,A}^2 \right) - \beta^{-1} \ln \frac{R_{\beta E_c}(n_{g,B})}{R_{\beta E_c}(n_{g,A})}.
\end{eqnarray}
The dissipated work, $W-\Delta F$, the central quantity in non-equilibrium fluctuation relations, can then be written
combining Eqs. (\ref{box7}) and (\ref{box10}) as
\begin{eqnarray} \label{box11}
W-\Delta F= 2 E_C \int_{n_{g,A}}^{n_{g,B}} \left(n_g - n\right) dn_g + \beta^{-1} \ln \frac{R_{\beta E_c}(n_{g,B})}{R_{\beta E_c}(n_{g,A})}.
\end{eqnarray}
In the basic gate protocol with $n_{g,A}=0$, $n_{g,B}=1$, used in the recent experiment by Saira et al. \cite{ops}, the expression for dissipated work assumes the form
\begin{eqnarray} \label{box12}
W-\Delta F= E_C(1-2\int_{0}^{1} n dn_g).
\end{eqnarray}
The same expression is obtained also if only states $n=0$ and $n=1$ are allowed in the calculation of the partition function. This is the appropriate choice for, e.g., experiments in the high charging energy regime $\beta E_C \gg 1$.

\section{First-principles derivation of the classical Hamiltonian}
\label{sec:lagrange}

We give an elementary derivation of Eq.~(\ref{box4}) based on Ref.~\cite{devoret}. Following the general procedure, the gate voltage source in Fig.~\ref{fig:box}(a) is represented by a large capacitance $C_S$ in which is stored intially a large charge $Q_S^0$ such that $Q_S^0/C_S = V_g$. First, one writes the Lagrangian in terms of the node fluxes $\phi_{a,b} := \int_{-\infty}^t \dot\phi_{a,b}(t') dt'$, where $\dot\phi_{a,b}$ refers to the potential of nodes $a, b$ as defined in Fig.~\ref{fig:box}(b). The result is
\begin{eqnarray} \label{eq:L1}
\mathcal{L}_\mathrm{full} = \frac{1}{2} C_j \dot\phi_a^2 + \frac{1}{2} C_g \left(\dot\phi_a - \dot\phi_b\right)^2 + \frac{1}{2} C_S \dot\phi_b^2.
\end{eqnarray}
It is important to distinguish between $\dot\phi_b$, referring to the variable voltage of node $b$, and $V_g$, the initial voltage of the source capacitor $C_S$. The generalized momenta are obtained with the usual relation $q_{a, b} = \frac{\partial\mathcal{L}_\mathrm{full}}{\partial \phi_{a,b}}$, and are equal to the sum of charges at the nodes $a, b$,
\begin{eqnarray}
q_a &= &C_j \dot\phi_a + C_g (\dot\phi_a - \dot\phi_b)\label{eq:L2a}\\
q_b &= &C_g (\dot\phi_b - \dot\phi_a) + C_S\dot\phi_b.\label{eq:L2b}
\end{eqnarray}
The Hamiltonian $H_\mathrm{full}$ is obtained as
\begin{eqnarray}
H_\mathrm{full} = \dot\phi_a q_a + \dot\phi_b q_b - \mathcal{L}_\mathrm{full}.
\end{eqnarray}
To obtain a reduced Hamiltonian describing the dynamics of the box charge exactly, one needs to subtract the energy of the capacitor $C_S$ from the full Hamiltonian. Subtracting the instantaneous energy $\frac{1}{2} C_S \dot\phi_b^2$, one obtains
\begin{eqnarray} \label{eq:LH0}
H_\mathrm{full} - \frac{1}{2} C_S \dot\phi_b^2 = \frac{1}{2}C_j\dot \phi_a^2+\frac{1}{2}C_g(\dot\phi_a-\dot\phi_b)^2,
\end{eqnarray}
which equals $H_0$ of Eq.~(\ref{box1}), as it should.
On the other hand, subtracting the energy $\frac{1}{2} C_S V_g^2$, which was stored initially in the capacitor should yield the Hamiltonian $H$ of Eq. (\ref{box3}) in the limit $C_S\rightarrow \infty$ \cite{devoret}. We obtain
\begin{eqnarray} \label{eq:LH1}
H_\mathrm{full} - \frac{1}{2} C_S V_g^2  = H_0 + \frac{1}{2} C_S(\dot\phi_b^2-V_g^2).
\end{eqnarray}
We next analyze the difference $\frac{1}{2} C_S(\dot\phi_b^2-V_g^2)$. The charge is conserved on the node $b$. Thus the gate charge at an arbitrary time  (assuming it is zero initially) reads $Q_g=C_S(V_g-\dot\phi_b)$. Inserting this charge conservation into the said difference yields
\begin{eqnarray}
\frac{1}{2} C_S (\dot\phi_b^2 - V_g^2) = -\frac{1}{2}Q_g(\dot\phi_b+V_g).
\end{eqnarray}
In the limit $C_S\rightarrow \infty$, we have $\dot\phi_b \rightarrow V_g$,
yielding
\begin{eqnarray} \label{eq:LH10}
H-H_0=\frac{1}{2} C_S (\dot\phi_b^2 - V_g^2) = -Q_gV_g
\end{eqnarray}
in accordance with Eq. (\ref{box3}).

\section{Comparison of the dissipated work $W-\Delta F$ and the dissipation $Q$ in tunneling}
\begin{figure}
    \includegraphics[width=13cm]{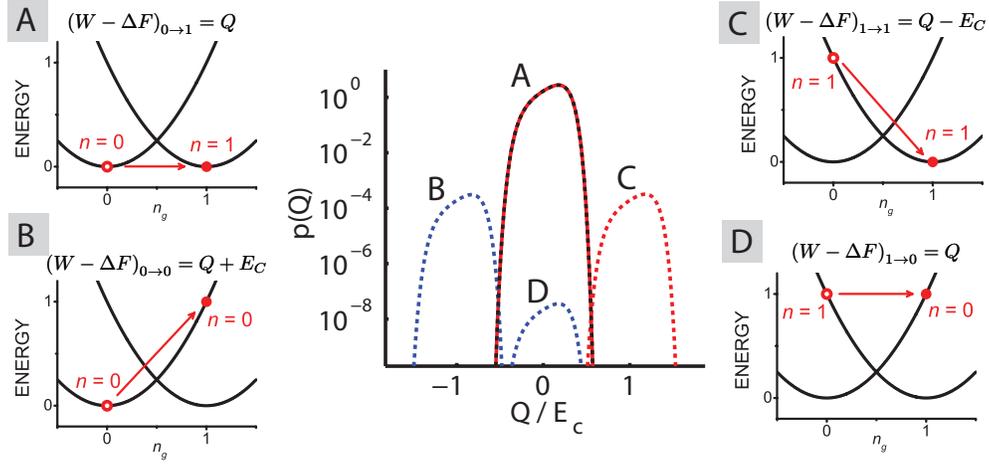}
    \caption{Graphical interpretation of Eqs. (\ref{box14}) - (\ref{box17}) to compare $W-\Delta F$ and $Q$ along gate trajectories $n_g:0\rightarrow 1$ with different outcomes. Outcome A, $n:0\rightarrow 1$, is the expected (and in the recent experiments~\cite{ops} the only) choice, where the system makes the expected transition. There $W-\Delta F=Q$. In the unlikely outcomes of B, C, and D, the electron numbers in the beginning and at the end are $(0,0)$, $(1,1)$, and $(1,0)$, respectively. The dashed lines in the middle panel give full simulated $Q$ distributions for different types of outcomes, where the capital letter refers to the four outcomes A - D, weighted with their probability. The simulations were made for a $f = 5~\mathrm{Hz}$ sinusoidal drive using the parameters of Ref.~\cite{ops}, $E_C/k_B =$ 1.94 K, temperature of the experiment $T = (k_B\beta)^{-1}= 214$ mK and the junction between superconducting (aluminium, $\Delta = 218$ $\mu$eV) and normal metal leads had a resistance $R_T = 100$ M$\Omega$.   The full distribution of $W - \Delta F$, the solid line, is indistinguishable from that of A of the dominating outcome. For the full distribution the horizontal axis is naturally replaced by $(W - \Delta F)/E_C$.}
    \label{wdfq}
\end{figure}
The dissipated energy in tunneling transitions along the gate drive was obtained in Ref. \cite{ap11} as
\begin{eqnarray} \label{box13}
Q = E_C\sum_i \pm (2n_{g,i}-1),
\end{eqnarray}
where the sum is over all back and forth tunneling events $i$ along the ramp, and $\pm$ refers to the direction of the jump: $+$ into the box, and $-$ out of the box.

We now compare the results (\ref{box12}) and (\ref{box13}) for different outcomes of the experiment. For convenience, we integrate Eq.~(\ref{box12}) by parts and obtain
\begin{eqnarray} \label{box12b}
W - \Delta F = E_C \left(1 - 2 n_f + 2 \sum \pm n_{g,i} \right) = E_C \left(1 - n_i - n_f \right) + Q.
\end{eqnarray}
where $n_{i(f)}$ is the charge state at the beginning (end) of the protocol and the sign in the summation is chosen as in Eq.~(\ref{box12}). For the most obvious case, i.e., the "successful" trajectories $n:0\rightarrow 1$ (meaning that the charge tunnels in the ramp as expected, perhaps with several intermediate back and forth transitions), we then have
\begin{eqnarray} \label{box14}
(W-\Delta F)_{0\rightarrow 1} = E_C\sum_i \pm (2n_{g,i}-1)=Q.
\end{eqnarray}
For trajectories $n:0\rightarrow 0$ (even number of jumps)
\begin{eqnarray} \label{box15}
(W-\Delta F)_{0\rightarrow 0}  = E_C[\sum_i \pm (2n_{g,i}-1)+1]=Q+E_C.
\end{eqnarray}
For the other types of trajectories we have
\begin{eqnarray} \label{box16}
(W-\Delta F)_{1\rightarrow 0} = E_C\sum_i \pm (2n_{g,i}-1)=Q
\end{eqnarray}
and
\begin{eqnarray} \label{box17}
(W-\Delta F)_{1\rightarrow 1}  = E_C[\sum_i \pm (2n_{g,i}-1)-1]=Q-E_C.
\end{eqnarray}
\section{Interpretation}
Figure \ref{wdfq} gives a graphical interpretation of the comparison in Eqs. (\ref{box14})-(\ref{box17}). The two quantities, the dissipated work, $W-\Delta F$, and dissipation $Q$ are directly related. The difference between them is that $Q$ gives the pure dissipation in the tunneling events, which is then released as heat typically to the phonon system, whereas $W-\Delta F$ is the sum of $Q$ and the energy stored in the electronic system in the gate ramp. In the properly designed experiment where the transition takes place in all realizations  ($\beta E_C \gg 1$) \cite{ops}, the two quantities are equal.

\section{Acknowledgements}
We thank Dmitri Averin, Tapio Ala-Nissila, Paolo Solinas and Aki Kutvonen for extensive discussions, and Angelo di Marco for careful proof-reading of the text. The work was supported by the Academy of Finland through its CoE program and by V\"ais\"al\"a foundation.

\end{document}